\documentclass{article}
\usepackage{setspace}
\usepackage{graphicx}
\usepackage{amsmath}
\usepackage{amssymb}
\usepackage[round]{natbib}
\usepackage{lscape}
\usepackage{bm}
\usepackage{comment}
\usepackage{multirow}
\usepackage{caption}
\usepackage[a4paper]{geometry}
\usepackage{booktabs}
\usepackage{xcolor}

\begin{document}
\title{Dirichlet Process Mixture Models for Regression Discontinuity Designs}
\author{Federico Ricciardi$^1$, Silvia Liverani$^{2,3}$, Gianluca Baio$^1$}
\date{    $^1$ Department of Statistical Sciences, University College London, London WC1E 6BT, UK\\%
    $^2$ School of Mathematical Sciences, Queen Mary University of London, London E1 4NS, UK\\%
$^3$ The Alan Turing Institute, The British Library, London NW1 2DB, UK \\%
}

\maketitle
\onehalfspacing

\begin{abstract}
The Regression Discontinuity Design (RDD) is a quasi-experimental design that estimates the causal effect of a treatment when its assignment is defined by a threshold value for a continuous assignment variable. The RDD assumes that subjects with measurements within a bandwidth around the threshold belong to a common population, so that the threshold can be seen as a randomising device assigning treatment to those falling just above the threshold and withholding it from those who fall just below. 

Bandwidth selection represents a compelling decision for the RDD analysis as the results may be highly sensitive to its choice. A number of methods to select the optimal bandwidth, mainly originating from the econometric literature, have been proposed. However, their use in practice is limited. 

We propose a methodology that, tackling the problem from an applied point of view, consider units' exchangeability, i.e., their similarity with respect to measured covariates, as the main criteria to select subjects for the analysis, irrespectively of their distance from the threshold. We carry out clustering on the sample using a Dirichlet process mixture model to identify balanced and homogeneous clusters. Our proposal exploits the posterior similarity matrix, which contains the pairwise probabilities that two observations are allocated to the same cluster in the MCMC sample. Thus we include in the RDD analysis only those clusters for which we have stronger evidence of exchangeability. 

We illustrate the validity of our methodology with both a simulated experiment and a motivating example on the effect of statins to lower cholesterol level, using UK primary care data.

%
%
%

\end{abstract}

\section{Introduction}
\label{sec:Intro}

The Regression Discontinuity Design (RDD) is a quasi-experimental design that estimates the causal effects of a treatment by exploiting the presence of a pre-determined treatment rule (either naturally occurring or regulated by on-going policies). The first publication on RDD was an application in education by \citep{ThistleCamp:1960}. Since then this framework has proved to be effective in a wide range of applications in other disciplines, including economics \citep{Cook:2008} and politics \citep{Lee:2008}. 
More recently there has been some interest in the RDD for epidemiology \citep{Boretal:2014,Deza:2015} and health and primary care applications \citep{LindenAdams:2012,OKeeffeetal:2014,Genelettietal:2015,Boretal:2017}.

The RDD can be applied in any context in which a particular treatment or intervention is administered according to a pre-specified rule linked to a continuous variable, referred to as the `assignment' or `forcing' variable: the treatment is then administered if the units' value for the assignment variable ($X$) lies above or below a certain threshold ($x_0$), depending on the nature of the treatment. 
If thresholds are strictly adhered to when assigning treatment, the design is termed \textit{sharp}, while when this is not the case it is termed \textit{fuzzy}.

The regression discontinuity design has become of particular interest in the definition of public health policies as it enables the use of routinely collected electronic medical records to evaluate the effects of drugs when these are prescribed according to well-defined decision rules. This is useful as government agencies such as the Food and Drug Administration (FDA) in the USA and the National Institute for Health and Care Excellence (NICE) in the UK are increasingly relying on guidelines for drug prescription in primary care. In fact we will use prescription of statins in the UK as our motivating example, but there is a wide range of potential applications including the prescription of anti-hypertensive drugs when systolic blood pressure exceeds 140mmHg or initiating antiretroviral therapy in patients with HIV-1 when their CD4 count has fallen to 350 cells/mm\textsuperscript{3} or below.



The RDD can mimic a randomised experiment around the threshold and the treatment effect at the threshold can be obtained averaging the outcomes in `small' bins in its proximity. 
The choice of the `bandwidth' is an important decision for an RDD analysis. because the results are highly sensitive to its choice, especially in all those cases in which the relationship between the assignment variable and the outcome, on both sides of the threshold, deviates from linearity.

In many applied studies \citep{Broockman:2009,Genelettietal:2015,li2015evaluating}, a standard strategy adopted to address the bandwidth issue is to produce local linear regression estimates obtained using data within a limited number of bandwidths (often not more than 3 or 4, sometimes defined with the guidance of experts in the field of study). 
Alternatively, more complex approaches can be adopted. 

Historically, these methods find their roots in the econometric literature and have close connection with the non-parametric estimation of the effect for RDD. Their common rationale is that the `optimal' bandwidth must be selected according to some criteria aimed to minimise an error term. The first proposal, by \citet{LudwigMiller:2005}, was based on a leave-one-out cross validation (CV) strategy in order to find the estimator minimising the mean integrated square error. Later, \citet{Imbens:2011} and \citet{CCT:2015} demonstrated that the CV method was a potential source of bias and that it was not reliable in any case when the design is fuzzy, and hence devised two slightly different minimisation methods based on the asymptotic mean square error. \citet{LeeLemieux:2009} give an overview of these approaches.

More recently, Local Randomization (LR) has been proposed by \citet{CattaFrandsen:2014} and used since in several applied papers \citep{skovron2015practical, li2015evaluating, MatteiMealli:2016} in an attempt to select a window around the threshold where the units can be seen as part of a randomised experiment. This approach, although motivated by a different intuition, shares a 
common trait with the other approaches outlined above (and further described in Section \ref{sec:LitRev}): they all aim at finding one bandwidth, having optimal properties under certain criteria and then use it within the RDD framework. As a consequence, they rely on what we named `all-or-nothing' selection mechanism: all units within the bandwidth are considered for the RDD analysis, but none of those outside.

In this paper, we propose an alternative approach to select the units to be included in a RDD analysis.
Similarly to the LR method, our approach originates from a pragmatic and applied point of view, focusing on units' exchangeability, an attribute rooted in the unconfoundness assumption
that guarantees that a RDD mimics a randomised control trial thanks to the similarity of the units above and below the threshold. 
However, our proposal has a more ambitious goal:
not only do we aim at including units for the RDD analysis based on their mutual similarity and not on their proximity to the threshold, but we also want to overcome the need of an `all-or-nothing' approach shared by all other methods existing in the literature. 

Our novel proposal is motivated by 
the idea that that units can be grouped in an unknown yet finite number of clusters in which the available covariates are balanced among units above and below the threshold. Using a Dirichlet process mixture model (DPMM), we cluster the units using continuous and categorical covariates to account for potential sources of confounding. By quantifying the internal similarity of the clusters obtained, only units belonging to the most homogeneous clusters are then used in the RDD analysis, irrespective of their distance from the threshold. Our proposal aims to a more effective sample selection, as it searches for `signal' in the data in farther regions from the threshold generally overlooked by the currently available bandwidth selection approaches and discards the `noise' from data points closer to the cut-off.



The paper is organised as follows. Section \ref{sec:Bayes} introduces the RDD and gives details about the Bayesian modelling framework we adopt for the analysis.
Section \ref{sec:LitRev} gives an overview of the current literature on bandwidth selection for regression discontinuity designs. 
Section \ref{sec:DPMM} presents the methodological core of the paper, where we discuss the use of clustering based on Dirichlet Process Mixture Models (DPMM) within the RDD framework and Section \ref{sec:clustRanSen} addresses the issue on cluster selection for the subsequent RDD analysis. Results on both a simulated experiment and a real dataset on the effect of statins on cholesterol level are given in Section \ref{sec:Res}. Finally a closing discussion is presented in Section \ref{sec:Disc}.

\section{Bayesian Inference for the Regression Discontinuity Design}
\label{sec:Bayes}

In this section we introduce the basic framework and notation for the RDD. 
Our work is motivated by an application of the regression discontinuity design to statin prescription in primary care. 
In the past years other works from our broader research group have originated from the same practical application and data, every time exploring a different aspect of the RDD \citep{OKeeffeetal:2014, Genelettietal:2015, Genelettietal:2016}.
In the UK, according to guidelines given by the National Institution for Health and Care Excellence (NICE), statins must be prescribed to patients whose 10-year risk score of developing a cardiovascular disease, 
predicted using a logistic regression model with a number of clinical and lifestyle indicators as independent variables, exceeds 20\% \citep{NICE}. This threshold has been revised in 2014, lowering it to 10\%, but we used pre-2014 data in this work and hence we applied the old cut-off value. 

Using the risk score as our forcing variable $X \in \{0,1\}$, a RDD analysis can assess whether statins treatment ($T$) can cause a reduction in Low-Density Lipoprotein (LDL) cholesterol (our outcome, $Y$), evaluated at the threshold set to $x_0=0.20$. To complete the basic notation, let $X^c=(X-x_0)$ be the centred assignment variable and $Z$ be the binary threshold indicator such that $Z=1$ if the forcing variable $X \ge x_0$ and $Z=0$ otherwise.
Note that $Z$ coincides with the treatment assignment variable $T$ when the design is sharp, but when RDD is applied to health and medical data it is reasonable to expect the design to be fuzzy, and hence the two variables not to coincide. In our motivating example this can be due both to GPs not adhering to NICE guidelines and to patients failing to take statins although prescribed to do so.  

It is widely known that the threshold indicator $Z$ is a special case of binary Instrumental Variable (IV) \citep{didelez2010}. For this reason, 
in order for the RDD analysis to be performed, a set of assumptions which can be derived from the IV literature must hold \citep{Hahnetal:2001, Genelettietal:2015}. 

While further theoretical and technical aspects of the RDD would add very little to the scope of this paper, being extensively covered in the literature 
\citep{ImbensLemieux:2008, vanderKlaauw:2008}, 
we make use of the next subsection to provide a more detailed overview of the Bayesian modelling framework we aim to use for the the estimation of the causal effect at the threshold.

\subsection{The causal effect}

 
Motivated by our example, where GPs' prescribing behaviour may not adhere to NICE guideline, our primary focus is on fuzzy designs, hence the effect we are interested in is the \textit{Local Average Treatment Effect} (LATE) at the threshold, defined as 
\begin{equation}
\mbox{LATE} = \dfrac{E(Y|Z=1) - E(Y|Z=0)}{E(T|Z=1) - E(T|Z=0)}. \nonumber
\end{equation}
The LATE numerator is equal to the \textit{Average Treatment Effect} (ATE).
The denominator, obtained as the difference in the expected treatment probabilities above and below the threshold, scales the ATE to account for the fuzziness of the design. 
In our motivating example, the LATE quantifies the change in LDL cholesterol at the 10-year risk threshold of 20\%.

More details about the assumptions that allow the identification of the above effect under a fuzzy observational regime can be found in \citet{NayiaAidan:2015}.

\subsubsection*{Models for the ATE}
Let the index $l \in \{a, b\}$ specify whether a unit's forcing variable value lies above or below the threshold. We decided to model the outcome, i.e., LDL cholesterol, separately for $l=a$ and $l=b$ as
\begin{gather}
y_{il} \sim N(\mu_{il}, \sigma^2);  \nonumber \\ 
\mu_{il} = \beta_{0l} + \beta_{1l} x^c_{il},  \nonumber
\end{gather}
where $x^c_{il}$ is the centred distance of variable $X$ from the threshold $x_0$ for the $i$-th individual belonging to $l$. 


In our examples in Section \ref{sec:Res}, both for the simulated scenarios and for the real data analysis, the relatively large sample size reduces the impact on posterior inference of distributional assumptions, especially for $\sigma$ which is likely dominated by information from observed data. With smaller samples or to ensure further robustness to prior on $\sigma$, other models are obviously possible, e.g., by considering an Half-Cauchy distribution \citep{polson:2012}. 

For the regression parameters, their prior distributions are chosen to reflect plausible LDL cholesterol levels for the observed range of risk scores. Prior specifications are defined as follows

%

\begin{align}
\beta_{0a} & = \beta_{0b} + \lambda;  \nonumber  \\
\beta_{0b} & \sim N(3.7; \sigma^2_{0b}=0.25);  \nonumber  \\
\beta_{1l} & \sim N(0; \sigma^2_{1l}=2);   \nonumber  \\
\sigma & \sim\mbox{Uniform}(0,5).  \nonumber 
\end{align}

To encode in the model some available information from the literature \citep{HTA} about the effect of statins in lowering cholesterol levels, we specify the prior distribution of $\lambda$ in order to be moderately informative, i.e.,
\begin{equation}
\lambda \sim N(-2,1).  \nonumber 
\end{equation}


Finally the ATE is calculated as  $\Delta_{\beta} = \beta_{0a}- \beta_{0b}$.




\subsubsection*{Models for the denominator of the LATE}
The total number of subjects treated on each side of the threshold is modelled, again separately for $l \in \{a, b\}$ as
\[
 \displaystyle\sum_{i=1}^{n_l} t_{il} \sim \mbox{Binomial}(n_l, \pi_l),
\]
where $n_l$ is the number of units either above or below the threshold.

Depending on the desired prior structure for $(\pi_b, \pi_a)$, we specify two models which, analogously to those in \citet{Genelettietal:2015}, have been named \textit{unconstrained} and \textit{flexible difference} model.

This means that for the unconstrained model we use vague Beta distributions
\[
\pi_l \sim \mbox{Beta}(1,1),
\]
with $l \in (a, b)$.

For the flexible difference model, we impose a mild prior structure acknowledging an actual difference between the treatment probabilities above and below the threshold, defining
\[
logit(\pi_a) \sim N(2, 1) \qquad \text{and} \qquad logit(\pi_b) \sim N(-2, 1).
\]
These distributions keep the bulk on the prior probability of treatment distributions, above and below the threshold, reasonably separate from one another, limiting the possibility that they result to be similar, while not constraining them to have a fixed difference.

The denominator for the LATE is then given by the difference
\[
\Delta_{\pi} = \pi_{a} - \pi_{b}.
\]

\medskip 

\begin{center}
  * * *
\end{center}


Depending on the chosen model for the denominator we get two different effects, i.e.,
\begin{equation}
\mbox{LATE}^{unct} = \dfrac{\Delta_{\beta}}{\Delta_{\pi}^{unct}} \qquad \text{and} \qquad \mbox{LATE}^{flex} = \dfrac{\Delta_{\beta}}{\Delta_{\pi}^{flex}} \nonumber
\end{equation}
for the unconstrained and flexible difference model respectively.

\section{A concise review of bandwidth selection methods}
\label{sec:LitRev}

In recent years there has been a surge in the interest of researchers for the choice of the bandwidth, as accounted by \cite{CattaneoVaz:2016} in their comprehensive review on the topic. In fact the definition of the bandwidth represents a fundamental decision for the RDD as there is both a clear 
link between the size of the bandwidth and the assumption of exchangeability and a trade-off with the precision of the estimates. If the bandwidth is small, units can be reasonably considered more similar to one another. 
If the bandwidth is too large, the converse is true, i.e., units could no longer be considered homogeneous.

In this section, we give an overview of the most prominent methods for neighbourhood selection in the literature.



\subsection{Cross Validation based approach}
\label{sec:CV}


The first approach found in the literature is based on a Cross Validation procedure as proposed in \citet{LudwigMiller:2005}\footnote{This is a working paper, later published as peer-reviewed article in a shortened version \citep{LudwigMiller:2007}}, also discussed 
by \citet{ImbensLemieux:2008}. 
Let 
\begin{equation}
\widehat{m}_h(X_i)=
\begin{cases}
\alpha_{a} + \beta_{a} X_i^c, & \text{if $X_i \ge x_0$,} \\
\alpha_{b} + \beta_{b} X_i^c, & \text{if $X_i<x_0$}   \nonumber
\end{cases}
\end{equation}
be the predicted value, using a bandwidth equal to $h$, of the outcome $Y$ regressed on the centred assignment variable $X_i^c$ when the $i$-th unit is left out from the calculation. The Cross Validation criterion is defined as:
\begin{equation}
\label{eq:CV}
CV_{Y,\delta}(h)= \dfrac{1}{N} \displaystyle\sum_{i:q_{X,\delta,b} \leq X_i \leq q_{X, 1-\delta,a}}^{N} \left( Y_i - \widehat{m}_h(X_i) \right)^{2}.
\end{equation}
Here $\widehat{m}_h(X_i)$ is estimated using only observations on one side of $X_i$
to mimic the fact that RDD estimates are based on regression estimates at the boundary. As a result, equation (\ref{eq:CV}) is an average of boundary prediction errors.
Furthermore $q_{X,\delta,b}$ and $q_{X,1-\delta,a}$ are the $\delta$-th and $(1-\delta)$-th quantiles of the empirical distribution of $X$ for the sub-samples `below' and `above' the threshold, respectively.
\citet{LudwigMiller:2007} suggest $\delta = 0.95$ to be appropriate, 
while \citet{ImbensLemieux:2008} and \citet{LeeLemieux:2009} state that $\delta=0.5$ represents a reasonable value, but the choice of an appropriate value varies according to the problem at hand and should be evaluated with care. 
The choice for the bandwidth given by this CV method is then represented by
\[
h_{CV}^{opt} = \arg\displaystyle\min_h CV_{Y,\delta}(h).
\]
This criterion leads to the bandwidth choice that minimises an approximation of the Mean Integrated Square Error (MISE): 
\[
\mbox{MISE}(h)= E \left[ \int_x \left( \widehat{m}_h(x) - m(x) \right) f(x) dx\right]
\]
where $m(x)=E[Y_i|X_i=x]$ and $f(x)$ is the density of the forcing variable.

In the case of a fuzzy RDD, \citet{ImbensLemieux:2008} suggest to use the smallest bandwidth selected by two $CV$ criteria applied separately to the outcome and to the treatment:
\[
h_{CV}^{opt} = \min \left( \arg\displaystyle\min_h CV_{Y,\delta}(h), \arg\displaystyle\min_h CV_{T,\delta}(h) \right),
\]
where $T$ denotes the treatment received and the formulation for $CV_{T,\delta}(h)$ is similar to that in (\ref{eq:CV}). 



\subsection{MSE expansion bandwidth selection}
\label{sec:MSE}


Both \citet{Imbens:2011} and \citet{CCT:2015} criticise the CV based approach, stating that this criterion relies on fitting the entire regression line between the $\delta$-quantile for the observation on the left and the ($1-\delta$)-quantile for those on the right, so that the result is not optimal for the problem at hand, being the aim of a RDD to estimate the effect at the threshold. 

Let $\widehat{\tau}$ be the estimated effect at the threshold for the RDD, the proposal of \citet{Imbens:2011} is based on minimising its asymptotic Mean Squared Error (MSE), 
i.e., $(\widehat{\tau} - \tau)^2$. Hence the MSE is defined as:
\begin{equation}
 \mbox{MSE}(h) = E[(\widehat{\tau} - \tau)^2] = E[((\widehat{\mu}_a - \mu_a)-(\widehat{\mu}_b - \mu_b))^2]
\end{equation}
where $\widehat{\mu}_b = \lim_{x \uparrow x_0} \widehat{m}_h(x)$ and $\widehat{\mu}_a = \lim_{x \downarrow x_0} \widehat{m}_h(x)$, i.e, the two regression estimators for the `true' models on the two sides of the threshold, i.e., $\mu_b = \lim_{x \uparrow x_0} m(x)$ and $\mu_a = \lim_{x \downarrow x_0} m(x)$.

To overcome some issues arising when trying to minimise the $\mbox{MSE}(h)$ directly, the authors use a first-order approximation around $h=0$ of the above quantity, which they term Asymptotic Mean Squared Error or $\mbox{AMSE}(h)$.  
The optimal bandwidth is therefore:
\begin{equation}
h_{IK} = \arg \min_h \mbox{AMSE}(h) = C_K \left(\dfrac{\sigma^2_a(x_0) + \sigma^2_b(x_0)}{f(x_0)(m_a^{\prime\prime}(x_0) + m_b^{\prime\prime}(x_0))^2} \right)^{1/5} N^{-1/5}
\end{equation}
where $C_K$ is a constant value depending on the choice of the kernel function $K(\cdot)$; $\sigma^2_b(x_0)$ and $\sigma^2_a(x_0)$ are the left and right limit at the threshold of the variance $\sigma^2(x) = Var(Y_i|X_i=x)$; $f(x)$ is the density of the forcing variable; $m_a^{\prime\prime}(x_0)$ and $m_b^{\prime\prime}(x_0)$ are the right and left limits of the second derivative of $m(x)=E[Y_i|X_i=x]$. 
The authors propose a data-dependent method to estimate $h_{IK}$ in three steps.

\citet{CCT:2015} considered that both previous methods produce bandwidths that are too wide, leading to confidence intervals 
with poor asymptotic coverage. The authors prove that correct asymptotic coverage is reached only if the bandwidth can satisfy the bias condition $nh_n^5 \rightarrow 0$, a requirement that none of the above mentioned methods can guarantee, 
leading to a first order bias in the distributional approximation. 
As a result, the conventional confidence intervals may substantially over-reject the null hypothesis of no treatment effect.

The authors propose a bias correction to address this problem that is able to improve the performance in finite samples. The final result is a generalisation of $h_{IK}$, which we term $h_{CCT}$, which allows for higher order polynomial to be used for the inference
and provides more robust confidence interval estimators.


\subsection{Local Randomization}
\label{sec:LocRAnd}
The Local Randomization (LR) approach selects a window around the cutoff in which the randomization assumption is likely to hold \citep{CattaFrandsen:2014,Sekhon:2016,CattaTiti:2016,Calonico:2016c}.

The rationale behind LR is that, because treatment assignment is assumed to be randomised by the threshold inside the window, the distribution of pre-intervention covariates should be the same for treated and untreated units. This observation is directly related
to the non-testable unconfoundness assumption needed for the RDD to infer valid causal estimators. For the RDD framework to be useful, the distribution of these covariates for treated and untreated units should be unaffected by the treatment $T$ within the bandwidth $h$ but should be affected by the treatment outside the window.

To find such desired bandwidth an iterative selection method is implemented. Starting from a arbitrary `small' bandwidth $\overline{h}_1$, for each one of the covariates, multiple tests of the null hypothesis of no effect of the treatment on the covariates is conducted and the
minimum p-value taken. 

If the minimum p-value obtained, $p_{1}$, is less than some pre-specified level the initial window was too large, hence one should decrease the initial window and start over. Otherwise, if $p_{1}$ is greater than the selected significance level, choose a larger window  $\overline{h}_2 \supseteq \overline{h}_1$, and go back to calculate a second iteration minimum p-value, $p_2$. The process continues until the minimum p-value is smaller than the desired level and a final bandwidth $h_{LR}$ is defined. 

\medskip 

\begin{center}
  * * *
\end{center}

The limited literature available and the lack of an unequivocal methodology for the bandwidth selection motivates our work: in the following we develop a more general RDD framework in which the choice of the bandwidth is not required, with positive effect on our results.

\section{Dirichlet Process Mixture Models}
\label{sec:DPMM}



In this paper, we propose a Dirichlet process mixture model to identify units that are similar (and so will be treated as exchangeable), above and below the threshold. We propose to identify these units by exploiting the characteristics of the clusters obtained with a Dirichlet process mixture model as described in the previous section. 

The Dirichlet process mixture model is a Bayesian nonparametric method for (unsupervised) clustering and applied in a variety of areas, such as retail analysis \citep{Pitkin:2019}, language processing and classification \citep{Crook:2009, Dreyer:2011, zhang:2005}, medical imaging \citep{dasilva:2007,Wachinger:2014}, epidemiology \citep{Hastie:2013,Molitor:2014,Pirani:2015,Mattei:2016,Liverani:2016,Coker:2016,Coker:2018} and genetics \citep{Papathomas:2012}.
The Dirichlet process was first introduced by \cite{Ferguson:1973} and is defined as a probability distribution over random probability measures. The distribution of a Dirichlet process is (almost surely) discrete, in that a random sample drawn from a Dirichlet process has a non zero probability that multiple draws will have identical values.  It is this discreteness property which makes the Dirichlet process ideal for clustering, as it avoids the need to determine the number of clusters a priori \citep{Neal:2000}. The basic Dirichlet process mixture model is formulated as follows:
\begin{eqnarray}
w_i|\theta_i&\sim&p(w_i|\theta_i)\\
\theta_i|G&\sim&G\\
G&\sim&DP(\alpha,G_0).
\end{eqnarray}

The Dirichlet process models the distribution from which data $w_1,\ldots, w_n$ are drawn as a mixture of distributions, $p(w_i|\theta_i)$,  where  each  parameter $\theta_i$ is  drawn  from  a  mixing distribution $G$ \citep{Neal:2000}.  $G_0$ is the base distribution, that is  the prior expectation of $G$, i.e., $E[G] =G_0$, and the concentration parameter $\alpha$ acts as an inverse variance where larger values of $\alpha$ result in smaller variances. 
Posterior inference from a DPMM utilises Markov chain Monte Carlo (MCMC) posterior simulation and our implementation uses the slice sampling procedure \citep{Kalli:2011}. Moreover, due to the nature of the stick-breaking construction of the Dirichlet process \citep{Sethuraman:1994}, label-switching moves are also implemented, to prevent the slice sampler from getting stuck in local modes \citep{Hastie:2015}. 

In  this  paper,  the  DPMM  is  implemented to model a mixture of Gaussian and categorical components. 
Let $S_i$ be the latent allocation variable so that if $S_i=c$ then individual $i$ is in cluster 
$c \in \{1, C\}$, then conditional on each cluster $c$, the likelihood for observable data $\mathbf{D}_i = (\mathbf{D}^1_i, \mathbf{D}^2_i)$ is 


%
\begin{equation}
p(\mathbf{D}_i|S_i=c,\Theta_{c}) = p(\mathbf{D}^1_i|\mu^{DP}_{c},\Sigma_{c})p(\mathbf{D}^2_i|\Phi_{c}) 
\end{equation}
where $\mathbf{D}^1_i = (D^1_{i,1}, ..., D^1_{i, J_1})$ is the subset of the $J_1$ continuous random variables in $\mathbf{D}_i$ and $\mathbf{D}^2_i= (D^2_{i,1}, ..., D^2_{i, J_2})$ is the subset of the $J_2$ categorical random variables in $\mathbf{D}_i$. Note that we are assuming independence between continuous and categorical data conditional on the cluster allocations. The cluster specific parameters are given by $\Theta_{c} = (\mu^{DP}_{c},\Sigma_{c},\Phi_{c})$, which are defined in detail below. 

For the continuous random variables, we have 
 
\begin{eqnarray}
p(\mathbf{D}^1_i|\mu_c^{DP},\Sigma_c)=(2\pi)^{-\frac{J_1}{2}}|\Sigma_c|^{-\frac{1}{2}}\exp\left\{-\frac{1}{2}(D^1_i-\mu_c^{DP})^\top \Sigma_c^{-1}(D^1_i-\mu_c^{DP})\right\}
\end{eqnarray}
and we choose $\mu_c^{DP}\sim\mathrm{Normal}(\mu_0^{DP},\Sigma_0)$ and $\Sigma_c\sim\mathrm{InvWishart}(R_0,\kappa_0)$ (for each $c$) for our prior model to obtain a conjugate model, permitting Gibbs updates for the parameters $\mu^{DP}$ and $\boldsymbol{\Sigma}$. 

For the discrete random variables, we have
\begin{eqnarray}
 p(\mathbf{D}^2_i|\Phi_c)=\prod_{j=1}^{J_2}\phi_{{S_i},j,X_{i,j}}. 
\end{eqnarray} 
For each individual $i$, $\mathbf{D}^2_i = (D^2_{1}, \ldots, D^2_{J_2})$ is a vector of $J_2$ locally independent discrete categorical random variables, where the number of categories for covariate $j=1,2,\ldots,J_2$ is $R_{j}$. 
Then we can write $\Phi_c=(\Phi_{c,1},\ldots,\Phi_{c,J_2})$ with $\Phi_{c,j} = (\phi_{c,j,1}, \phi_{c,j,2}, \ldots, \phi_{c,j,R_j})$.
Letting $\mathbf{a}=(\mathbf{a_1}, \mathbf{a_2}, \ldots, \mathbf{a_{J_2}})$, where 
$a_{j}=(a_{j,1},\ldots,a_{j,R_{j}})$ 
and adopting conjugate Dirichlet priors $\Phi_{c,j} \sim \mathrm{Dirichlet}(a_{j})$, each $\Phi_{c,j}$ can be updated directly using Gibbs iterations. 

As each iteration of the MCMC Gibbs sampler provides an estimate of the cluster labels, Partitioning Around Medoids (PAM) was used to obtain an overall estimate of the optimal number of clusters \citep{KaufmanRousseeuw:2005}. As the number of clusters varies between iterations, the proposed method uses the posterior 
similarity matrix $\mathbf{P}$. 
The  best  clustering  is selected by maximising an associated clustering score \citep{Molitor:2010}.
The Dirichlet process mixture model described above is available in the R package PReMiuM \citep{Liveranietal:2015}.

\section{Cluster Ranking and Selection }
\label{sec:clustRanSen}

Once we have identified units that are similar to one another using a Dirichlet process mixture model, above and below the threshold, we must identify the most suitable clusters for the RDD analysis. We propose to identify clusters that are \textit{balanced} and \textit{homogeneous}. These concepts have been extensively exploited in several branches of statistics, most notably by the Propensity Score Weighting literature \citep{Crump:2009, Li:2018}, where overlap in covariates between treatment groups is a desired feature to estimate average treatment effects for sub-populations defined according to the propensity score.
 

A cluster is \textit{balanced} when it has enough units on both sides of the threshold. As many small clusters are usually fully above, or below, the threshold, it is important to ensure that we consider balanced clusters for the RDD analysis. 
Therefore, we define a value $\pi^{Z}$ for the proportion of units with assignment value above the threshold, i.e., $X_i \ge x_0$ for $i \in K_c$.
We call $\pi^{Z}_c$ the proportion of units in cluster $c$ with $Z_i =1$, i.e., for which the assignment variable is greater than the threshold $x_0$.
We the empirically set a constant parameter $\zeta$
 , deeming a cluster \textit{balanced} if the proportion $\pi^{Z}_{c}$ falls within an acceptable range, i.e., $\dfrac{1}{\zeta} \leq \pi^{Z}_{c} \leq \dfrac{\zeta-1}{\zeta}$. We then discard unbalanced clusters, leaving us with $C' \leq C$ clusters. 

A cluster is \textit{homogeneous} (or compact) when the observations within it are very similar to one another. 
However, modelling with a mixture model does not always result in clusters of similar observations. For example, a Gaussian mixture model with a fully flexible covariance matrix may incur in large within-cluster dissimilarities compared to a model in which covariance matrices are assumed to be equal or spherical: observations that are modelled well by a common probability distribution are not necessarily close. For example, in the case of a 2-dimensional Gaussian distribution with a high correlation, the maximum distance between the further observations can be significant. Generally, the mixture model does not come with implicit conditions that ensure the separation of clusters \citep{Henning:2013}. Therefore, we employ the Dirichlet process mixture model to exploit its flexibility, but we must take a close look to the homogeneity of each cluster.

We propose to rank clusters based on their homogeneity. The concept of homogeneity is widely explored in the clustering literature \citep{Everitt:2011} and relies on the idea that if properly identified, units in a cluster must have a cohesive structure. The most straightforward way to formalise that all objects within a cluster should be similar to each other is the average within-cluster distance, a commonly used index for cluster internal validation \citep{Henning:2017arX}. We employ a version of this within-cluster index based on the posterior  
similarity matrix $\mathbf{P}$ obtained post-processing the output of the Dirichlet process mixture model. The values in $\mathbf{P}$ are the pairwise probabilities that two observations are allocated to the same clusters in the MCMC sample. 
As such, adapting the definition of dissimilarity from \cite{Henning:2017arX}, 
we can define a 
similarity function $s: \mathcal{V}^2 \longmapsto \mathbb{R}^+_0$ so that $s(v_1, v_2) = s(v_2, v_1)\ge 0$ and $s(v_1, v_1) = 1$, 
where $v_1$ and $v_2 $ are elements from the space $\mathcal{V}$, i.e., the objects needed to be clustered. 
This similarity function can be used to compute the within-cluster homogeneity. 

For $c' = 1, \ldots, C'$ 
let $\mathcal{K'}=\{K_1, \ldots, K_{C'} \}$ with $K_{c'} \subseteq \mathcal{V}$ be the clustering set where the unbalanced clusters have been removed so that $n_{c'} = \vert K_{c'} \vert$ is the number of units in cluster $c'$. 
Let $p_{l,v}$ be the elements of the similarity matrix $\mathbf{P}$. For each cluster this within-cluster homogeneity index can be calculated as:
\begin{equation}
   I_{c'} = \dfrac{2}{n_{c'}(n_{c'}-1)} \sum_{l=1}^{n_{c'}} \sum_{v \leq l}^{n_{c'}} p_{l,v}.
\end{equation}
A lower within-cluster index is an indicator of a more homogeneous cluster, with 0 being the minimum value for $I_{c'}$. We exploit this measure of homogeneity to rank the clusters from the least homogeneous to the most homogeneous. We relabel the index as $I_{({c'})}$ for ${c'}=1,\ldots,C'$ such that $I_{(1)} < I_{(2)} < \ldots < I_{(C')}$. 


Among the balanced clusters, we propose to use homogeneity to select the clusters to include in our model. We propose the following four criteria. 

\begin{enumerate}

\item We include clusters until the relative difference between the homogeneity for the $c'$-th and ($c'+1$)-th ordered clusters is within a 10\% margin, that is, all ordered clusters from 1 to ${c'}$ such that 
\begin{equation*}
\frac{I_{({c'}+1)}-I_{({c'})}}{I_{({c'})}} < 0.10
\end{equation*}
for ${c'}=1,\ldots, C'$. We refer to this criteria as \textit{inc10}.

\item We include the first quartile of the balanced clusters, that is, all clusters ${c'}$ with
\begin{equation*}
I_{(\left\lceil h \right\rceil)} \quad \mbox{such that} \quad h \le C'/4.
\end{equation*}
We refer to this method as \textit{c25}.

\item We include clusters starting from the most homogeneous until the sample includes at least half of the units from the entire cohort, that is, all clusters $c'$ with $c'=1,\ldots, C'$ such that 
\begin{equation*}
\sum_{c'=1}^{C'-1} n_{({c'})}< N/2 \quad \mbox{and} \quad \sum_{{c'}=1}^{C'} n_{({c'})} \ge N/2
\end{equation*}
where $n_{(c')} $ is the cardinality of the $c'$-th cluster, ordered according to the homogeneity index $I_{(s)}$.
We refer to this criteria as \textit{n50}.

\item We named to this final criteria as \textit{n25} as it is similar to \textit{n50}, but only considering 
one quarter of the units from the entire cohort, that is, all clusters $c'$ with $c'=1,\ldots, C'$ such that 
\begin{equation*}
\sum_{c'=1}^{C'-1} n_{(c')}< N/4 \quad \mbox{and} \quad \sum_{c'=1}^{C'} n_{(c')} \ge N/4.
\end{equation*}
\end{enumerate}

The four strategies detailed above define four (possibly) different sub-samples of the partition obtained applying a Dirichlet process mixture model as in Section \ref{sec:DPMM}. RDD analysis, as detailed in Section \ref{sec:Bayes}, is hence performed for each of the sub-samples of units irrespective of their distance from the threshold ($x_0$).

\section{Applications and results}
\label{sec:Res}


We make use of our methodology for an application to primary care prescription: according to the guidelines given by the National Institute for Health and Care Excellence (NICE) between 2008 and 2014, statins should have been prescribed in the UK to patients with 10-year cardiovascular disease (CVD) risk scores, calculated via the so called Framingham Risk Score \citep{DAgostinoetal:2008}, in excess of 20\%. To illustrate our methodology and check its performance, we use 
statins prescriptions data from The Health Improvement Network (THIN - \texttt{www.the-health-improvement-network.com}) a large primary care database that provides anonymised longitudinal general practice data on patients' diagnostic and prescribing records from more than 500 general practices across the UK. The database is broadly representative of the UK population \citep{bourke2004feasibility}.

In the following Sections we will present results obtained using our methodology both on a realistically simulated dataset (Section 
\ref{sec:Simul}) and on a subset of data from THIN patients (Section \ref{sec:Real}). The simulated experiment makes a formal comparison between different methods (i.e., our DPMM clustering based approach and other relevant bandwidth selection criteria), while the real-data application showcases 
how our methodology can be useful in practice. 
In both cases, the values of three key covariates are used to cluster units with our DPMM: age, systolic blood pressure and high-density lipoprotein (HDL) cholesterol. 
With the same data, we have obtained results of RDD analyses using established bandwidth selection methods: those based on MSE (i.e., IK and CCT) and Local Randomization (LR) as detailed in Sections \ref{sec:MSE} and \ref{sec:LocRAnd} as well as two arbitrarily selected windows, i.e., bandwidth of width 0.05 and 0.1 on each side of the threshold. Appropriate functions from R packages \texttt{rdd}, \texttt{rdrobust} and \texttt{rdlocrand} are used to estimate $h_{IK}$, $h_{CCT}$ and $h_{LR}$ respectively.

\subsection{Simulated example}
\label{sec:Simul}

For this example we followed the same approach as \citet{Genelettietal:2015} and used simulated data originated from the THIN database (details about the simulation algorithm can be found on the supplementary material of that paper). In particular data are obtained under a simulation scenario in which the risk score is a strong instrument for the treatment, the treatment effect size is equal to -2 and there is low level confounding. 
Both statins treatment status and the LDL cholesterol outcome are simulated to mimic realistic values.  

We have simulated 100 datasets with $N=5,720$ units
 and for each of them, separately, we clustered the units using the DPMM approach.
Then we selected the most homogeneous clusters based on the four criteria detailed in Section \ref{sec:clustRanSen}. The range for acceptable assignment probability for each cluster is $\dfrac{1}{10} \leq \pi^Z_c \leq \dfrac{9}{10}$, i.e., $\zeta=9$. 
This is done to account for the fact that in most of the clusters the assignment probabilities are not very well balanced (i.e., to prevent a too drastic exclusion of ineligible clusters). 
Finally we performed RDD Bayesian analysis and combined the results to obtain $\mbox{LATE}^{unct}$ and $\mbox{LATE}^{flex}$. 


%

\begin{figure}[h!]
\centering
\includegraphics[scale=0.50]{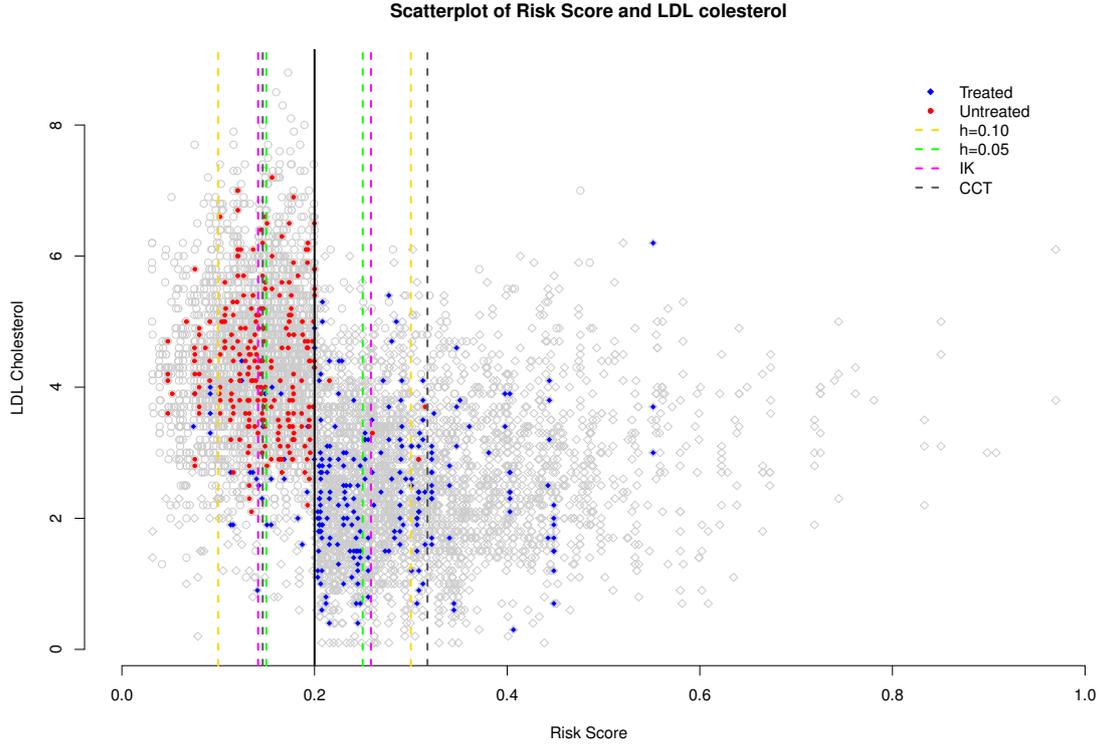}
\caption{\label{fig:subsetS_dist} Scatterplot of 10-year CVD risk score vs. LDL cholesterol for one of the realistically simulated datasets, highlighting the units selected for the RDD analysis using the 
`c25' strategy (treated (blue) and untreated (red)), compared with other bandwidth selection methods (LR bandwidths are not depicted as they are too close to the threshold line).} 
\end{figure}

Figure \ref{fig:subsetS_dist} 
 gives a visual representation of how units are selected according to different bandwidth methods compared with the our DPMM framework combined with the c25 criteria: solid red dots and blue diamonds represent the selected units out of the whole initial sample, represented using void grey markers. Vertical lines show the bandwidths selected with some of the methods described in Section \ref{sec:LitRev}. Note that LR bandwidths are not shown to avoid confusion, as they are too close to the threshold. 

\begin{table}[h!]
\caption{Results for the simulated example.}
\centering
\label{tab:resDist_SD}
\begin{tabular}{rlrrrr}
  \hline
 & method & MEDIAN & MEAN & LOWER & UPPER \\ 
  \hline
  $\mbox{LATE}^{flex}$ & \multirow{2}{*}{\parbox{1.8cm}{inc10}} & -2.16 & -2.19 & -3.10 & -1.47 \\ 
  $\mbox{LATE}^{unct}$ &  & -2.26 & -2.42 & -3.55 & -1.41 \\ 
    \hline
  $\mbox{LATE}^{flex}$ & \multirow{2}{*}{\parbox{1.8cm}{c25}} & -1.96 & -1.97 & -2.28 & -1.66 \\ 
  $\mbox{LATE}^{unct}$ &  & -1.97 & -1.97 & -2.28 & -1.67 \\ 
    \hline
  $\mbox{LATE}^{flex}$ & \multirow{2}{*}{\parbox{1.8cm}{n50}} & -2.03 & -2.03 & -2.18 & -1.89 \\ 
  $\mbox{LATE}^{unct}$ &  & -2.03 & -2.03 & -2.18 & -1.89 \\
  \hline
  $\mbox{LATE}^{flex}$ & \multirow{2}{*}{\parbox{1.8cm}{n25}} &  -1.96 & -1.96 & -2.16 & -1.77 \\
  $\mbox{LATE}^{unct}$ &  & -1.96 & -1.96 & -2.16 & -1.77 \\  
   \hline
  $\mbox{LATE}^{flex}$ & \multirow{2}{*}{\parbox{1.8cm}{LR}}& -1.44 & -1.46 & -2.79 & -0.27 \\ 
  $\mbox{LATE}^{unct}$ & & -1.54 & -1.62 & -3.52 & -0.27 \\ 
   \hline
  $\mbox{LATE}^{flex}$ & \multirow{2}{*}{\parbox{1.8cm}{CCT}}& -2.05 & -2.05 & -2.21 & -1.90 \\ 
  $\mbox{LATE}^{unct}$ & & -2.05 & -2.05 & -2.21 & -1.90 \\ 
   \hline
  $\mbox{LATE}^{flex}$ & \multirow{2}{*}{\parbox{1.8cm}{IK}}& -2.08 & -2.08 & -2.24 & -1.93 \\ 
  $\mbox{LATE}^{unct}$ & & -2.08 & -2.09 & -2.24 & -1.93 \\ 
   \hline
  $\mbox{LATE}^{flex}$ & \multirow{2}{*}{\parbox{1.8cm}{$h=0.10$}}& -2.10 & -2.10 & -2.25 & -1.94 \\ 
  $\mbox{LATE}^{unct}$ & &  -2.10 & -2.10 & -2.25 & -1.95 \\ 
   \hline
  $\mbox{LATE}^{flex}$ & \multirow{2}{*}{\parbox{1.8cm}{$h=0.05$}}& -2.10 & -2.10 & -2.27 & -1.93 \\ 
  $\mbox{LATE}^{unct}$ &&  -2.10 & -2.10 & -2.27 & -1.93 \\ 
   \hline
\end{tabular}
\end{table}

\begin{figure}[h!]
\centering
\includegraphics[scale=0.47]{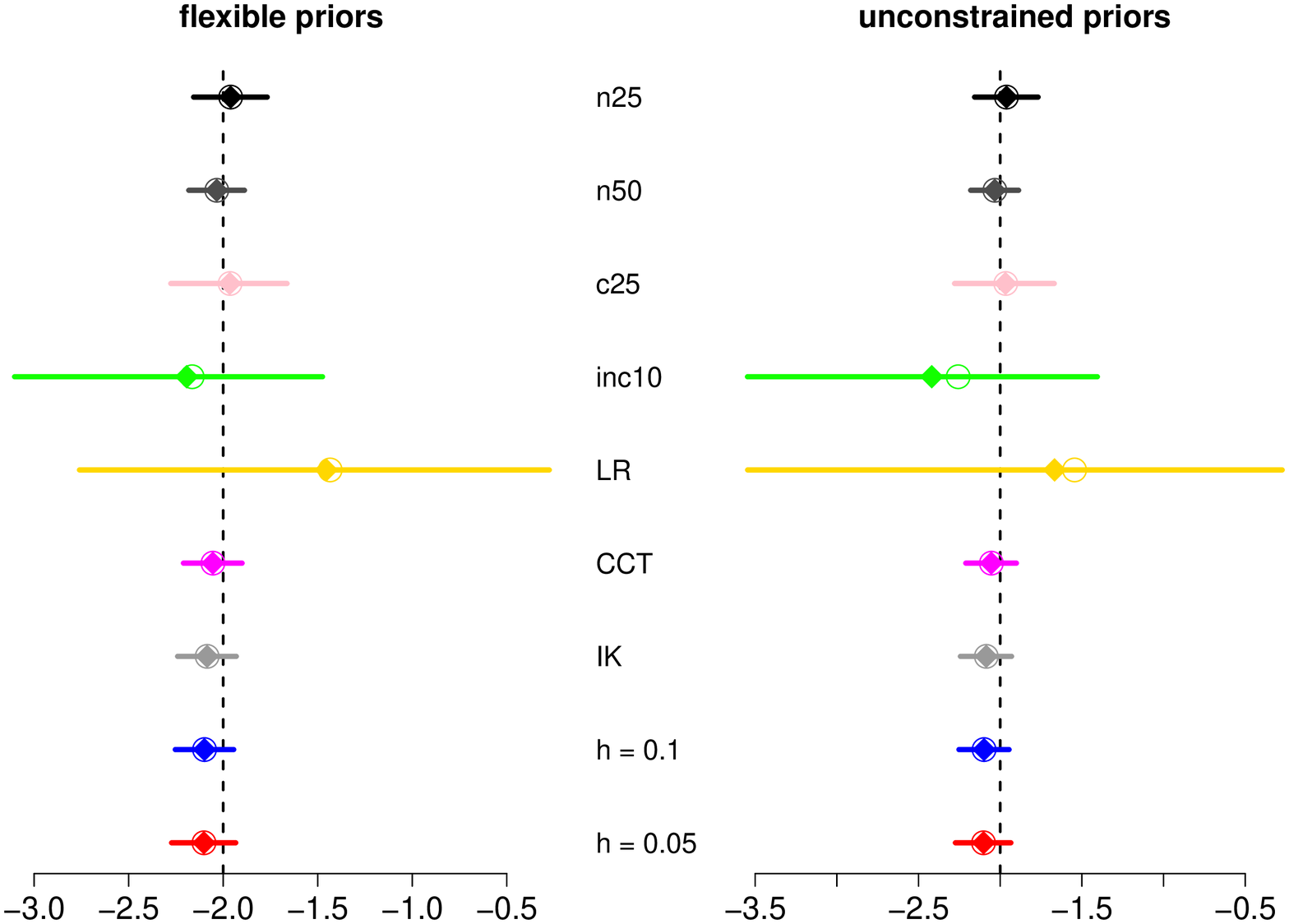}
\caption{\label{fig:compresS_dist} Comparison of results for the simulated example.} 
\end{figure}

Table \ref{tab:resDist_SD}
  and Figure \ref{fig:compresS_dist} show the results of these scenarios. 
It is worth noticing that flexible and unconstrained estimators give very similar results. 
Among the four cluster selection strategies we propose, \textit{c25}, \textit{n50} and \textit{n25} all show a reduced bias than those obtained using other established methods - i.e. CCT, IK, LR and arbitrarily-selected fixed-width bandwidths - while results for strategy \textit{inc10} are considerably less reliable.
Precision of all estimators is comparable for all strategies but \textit{inc10}, which shows wider credible intervals.

\subsection{Real data - Statins prescription in the UK.}
\label{sec:Real}

In this second example, we considered a subset of patients from THIN: male individuals aged from 50 to 70 who had not previously received a statin prescription nor suffered from  a CVD event and for whom the Framingham risk score was recorded by the GP during the time between 1 January 2007 and 31 December 2008. 
We further restricted the analysis to non-diabetic and non-smoking patients, so that the total number of units is 1386.

\begin{figure}[h!]
\centering
\includegraphics[scale=0.40]{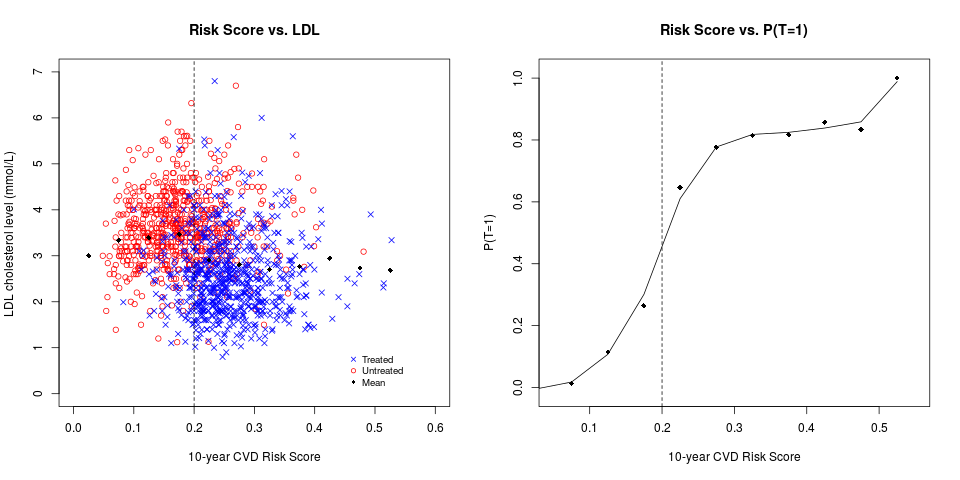}
\caption{\label{fig:DiagnosticReal} The left-hand plot shows 10-year CVD risk score vs. LDL cholesterol for treated (blue) and untreated (red), and the mean cholesterol lever within some equally spaced bins (black); the right-hand side plot shows risk score vs. the estimated probability of treatment, within the same bins. The dashed line indicates the threshold of 0.2.} 
\end{figure}

Figure \ref{fig:DiagnosticReal} shows why we believe a RDD is appropriate for the data at hand, clearly highlighting a discontinuity at the threshold for both the LDL level and the probability of statins prescription. Also we can appreciate the substantial fuzziness of the data, so that the LATE estimators are appropriate in this setting.

\begin{figure}[h!]
\centering
\includegraphics[scale=0.50]{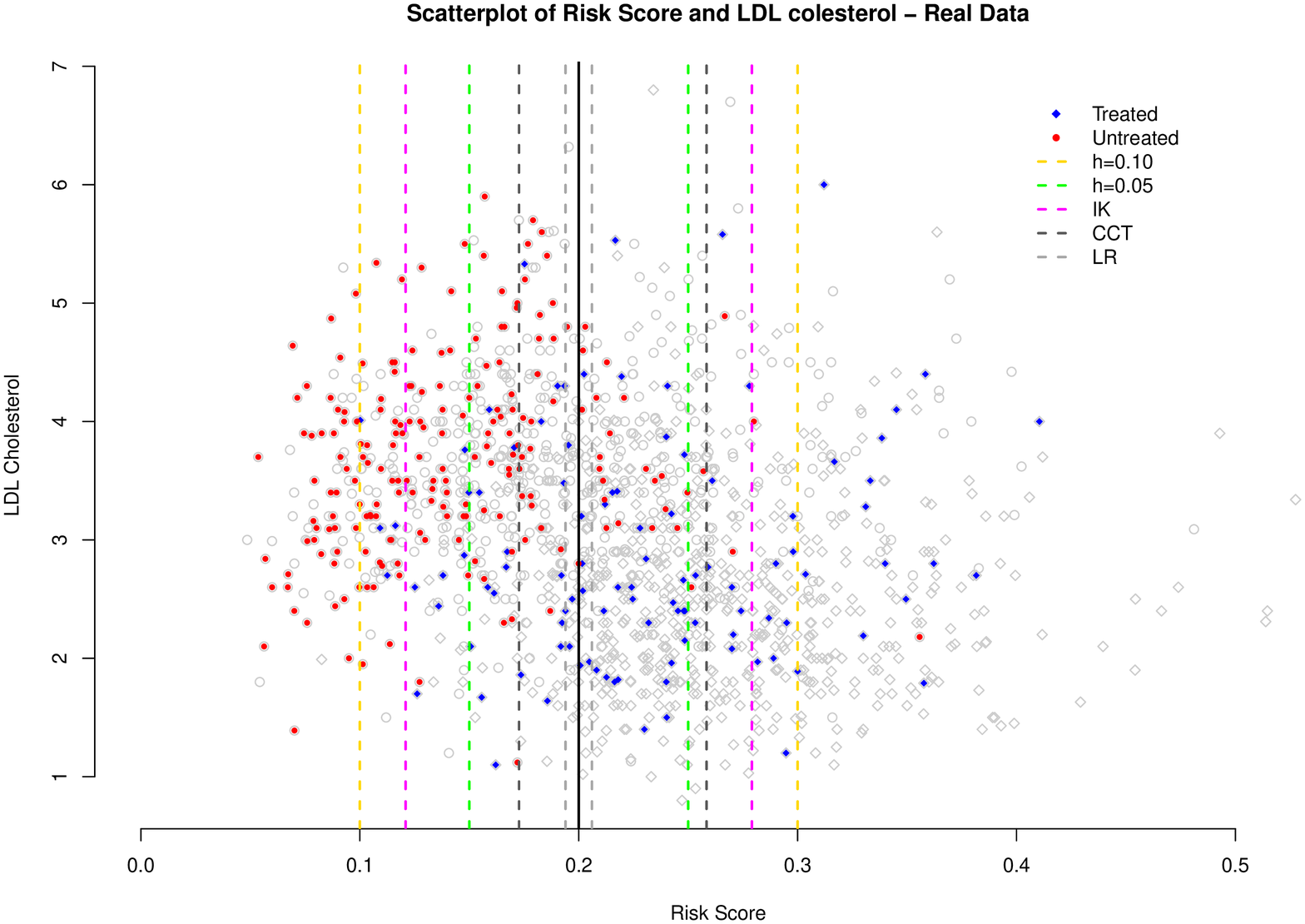}
\caption{\label{fig:subsetREAL_dist} Scatterplot of 10-year CVD risk score vs. LDL cholesterol for Real case, highlighting the units selected for the RDD analysis using 
the `c25' strategy (treated (blue) and untreated (red)), compared with other bandwidth selection methods.} 
\end{figure}

\begin{table}[h!]
\caption{Results for example based on real data.}
\centering
\label{tab:resDist_real}
\begin{tabular}{rlrrrr}
  \hline
 & method & MEDIAN & MEAN & LOWER & UPPER \\ 
  \hline
  $\mbox{LATE}^{flex}$ & \multirow{2}{*}{\parbox{1.8cm}{inc10}} & -1.01 & -1.03 & -1.58 & -0.57 \\ 
  $\mbox{LATE}^{unct}$ &  & -1.10 & -1.10 & -1.74 & -0.49 \\ 
  \hline
  $\mbox{LATE}^{flex}$ & \multirow{2}{*}{\parbox{1.8cm}{c25}} & -1.02 & -1.04 & -1.59 & -0.55 \\ 
  $\mbox{LATE}^{unct}$ &  & -1.09 & -1.09 & -1.67 & -0.49 \\ 
  \hline
  $\mbox{LATE}^{flex}$ & \multirow{2}{*}{\parbox{1.8cm}{n50}} & -0.95 & -0.96 & -1.32 & -0.67 \\ 
  $\mbox{LATE}^{unct}$ &  &  -0.97 & -0.97 & -1.30 & -0.68 \\ 
  \hline
  $\mbox{LATE}^{flex}$ & \multirow{2}{*}{\parbox{1.8cm}{n25}} & -1.12 & -1.12 & -1.49 & -0.76 \\
  $\mbox{LATE}^{unct}$ &  & -1.14 & -1.14 & -1.56 & -0.71 \\ 
   \hline
  $\mbox{LATE}^{flex}$ & \multirow{2}{*}{\parbox{1.8cm}{LR}} & -2.07 & 0.53 & -21.36 & 26.04 \\ 
  $\mbox{LATE}^{unct}$ & &  -1.79 & 3.52 & -28.67 & 57.38 \\ 
   \hline
  $\mbox{LATE}^{flex}$ & \multirow{2}{*}{\parbox{1.8cm}{CCT}} & -1.53 & -1.56 & -2.21 & -0.95 \\ 
  $\mbox{LATE}^{unct}$ & & -1.55 & -1.58 & -2.44 & -0.94 \\ 
   \hline
  $\mbox{LATE}^{flex}$ & \multirow{2}{*}{\parbox{1.8cm}{IK}} & -1.17 & -1.18 & -1.63 & -0.82 \\ 
  $\mbox{LATE}^{unct}$ & & -1.19 & -1.19 & -1.61 & -0.83  \\
     \hline
  $\mbox{LATE}^{flex}$ & \multirow{2}{*}{\parbox{1.8cm}{$h=0.10$}} & -1.04 & -1.05 & -1.40 & -0.74 \\ 
  $\mbox{LATE}^{unct}$ & & -1.09 & -1.09 & -1.42 & -0.72 \\
     \hline
  $\mbox{LATE}^{flex}$ & \multirow{2}{*}{\parbox{1.8cm}{$h=0.05$}} & -1.39 & -1.40 & -1.98 & -0.94  \\ 
  $\mbox{LATE}^{unct}$ & & -1.42 & -1.43 & -2.00 & -0.99  \\ 
   \hline
\end{tabular}
\end{table}

\begin{figure}[h!]
\centering
\includegraphics[scale=0.47]{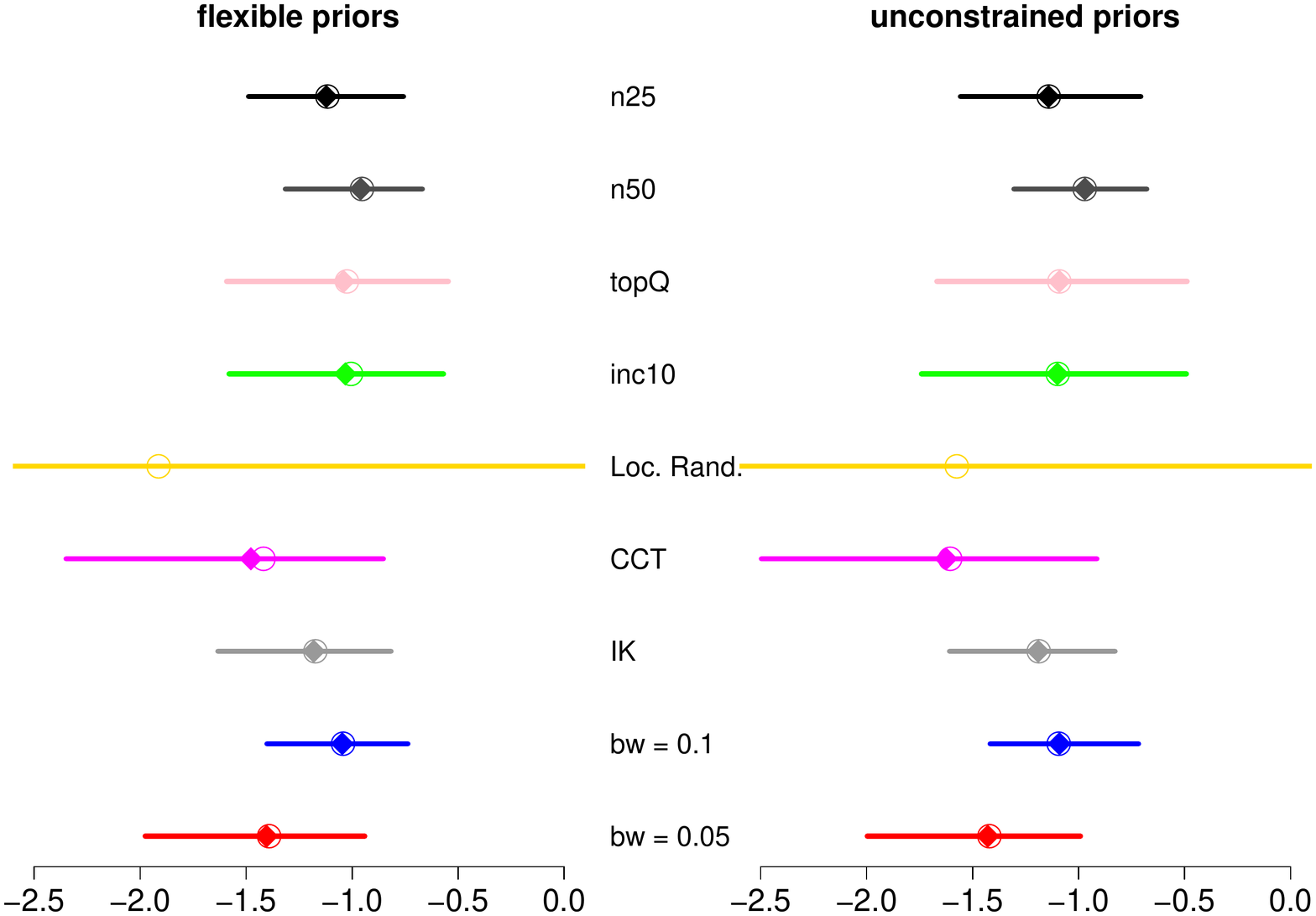}
\caption{\label{fig:compresREAL_dist} Comparison of results for the Real Case.} 
\end{figure}

For the clustering selection process, the range for acceptable assignment probability for each cluster is set to $\dfrac{1}{10} \leq \pi^Z_c \leq \dfrac{9}{10}$, i.e., $\zeta=9$. Table \ref{tab:resDist_real} and Figure \ref{fig:compresREAL_dist} show the results. Obviously there is no real value to compare the results of the estimators with, but there are a few aspects of interest nonetheless. All our DPMM estimators, including \textit{inc10}, produce similar results, irrespective of which cluster selection method is used, with \textit{n50} and \textit{n25} strategies both producing more precise estimates. It is also interesting to note how, in this case, the LR method produces very wide credible intervals for both $\mbox{LATE}^{unct}$ and $\mbox{LATE}^{flex}$, as this method is not able to pick a large enough subset of similar unit, being constrained to limit the search within nested windows. Results from both MSE based and arbitrary-selected bandwidths appear substantially different: CCT estimators are less precise that the DPMM based ones, and only $h=0.10$ produces results similar, in median, to those obtained applying our DPMM and cluster selection.

\section{Conclusions}
\label{sec:Disc}

We have proposed a novel, data-driven approach to deal with the 
bandwidth selection issue for the regression discontinuity design from a different perspective than those adopted in  the available literature.
Our approach originates from the idea that what matters the most in a regression discontinuity design is the exchangeability of the units included in the analysis, i.e., their homogeneity with respect to know observable characteristics. Following this rationale, it is reasonable to believe that subgroups of units might  share common characteristics irrespective of their distance from the threshold. This, we believe, represents the 
most appealing aspect of this framework: 
instead of relying on the `all-or-nothing' approach, which is implicit with any of the 
currently available bandwidth selection methodologies, we propose a tool that is capable of using all the available information from all the individuals showing homogeneous covariates and balanced forcing variable. 

Furthermore, when compared with the Local Randomization method which is similar in principle to ours, our DPMM clustering approach has the merit of tackling exchangeability more directly: while the former tests the null hypothesis of no effect of the treatment on each observed confounder separately in a univariate way, the latter, relying on clustering methods, evaluates the homogeneity of the considered covariates in a joint, more comprehensive approach.


The results of the RDD analysis using our DPMM clustering framework, especially in combination with \textit{c25} cluster selection strategy, compared favourably in terms of bias with those obtained following other bandwidth selection approaches, i.e., CCT and IK (methods that are specifically designed to minimise the bias of the causal estimator ), LR and with the arbitrarily-selected fixed-width bandwidths. 

We are aware of the limitations to our approach. In particular we acknowledge the issue that, due to the complexity of the DPMM which involves the estimation of a latent clustering structure, our analysis is more time consuming than those based on other bandwidth selection methods, an issue that is amplified as the number of clustering covariates increases. Due to label switching and the lack of a specific parameter to target, it is also hard to assess Bayesian DPMM convergence with the usual MCMC diagnostics. We remain convinced that these limitations are a reasonable price to pay in order to be able to overcome the `all-or-nothing' bandwidth approach.
A further limitation is the fact that our method relies on the availability of observed data or known confounders, although this issue is not exclusive of our approach as it is shared with Local Randomization method as well.

As a final remark, we think it is useful to note that we are not advocating the indiscriminate use of our methodology in any given RDD analysis. Expert assessment of any application and a proper evaluation of the plausibility of the RDD assumptions must always constitute the ground for subsequent analyses. Also, availability of covariates data and their role as potential confounders must be assessed beforehand. 
A degree of subjectivity remains in the choice of parameter $\zeta$, for which an assessment of the balance of the forcing variable has been proposed as a way to deal with clusters with unbalanced representation on both sides of the threshold, but the magnitude of the reasonably allowed unbalance represents an application-specific feature and it is left for the practitioner to be determined.

Rather than being a `one-size-fits-all' tool, our proposal offers an alternative approach to identify the units to be included in the RDD analysis in a more targeted way than the bandwidth selection methods currently available. Thoughtful use of our proposed DPMM clustering framework can prove valuable in all those RDD applications where exchangeability is regarded as a key feature and where traditional methods do not offer viable solutions to tackle it.

%
%
%
%

\section*{Acknowledgements}
This research has been funded by a UK MRC grant MR/K014838/1. Approval for this study was obtained from the Scientific Review Committee of the THIN in August 2014.

\bibliographystyle{plainnat}
\bibliography{rdd}
\end{document}